\documentclass[twoside,twocolumn,9pt]{article}
\usepackage{extsizes}
\usepackage[super,sort&compress,comma]{natbib} 
\usepackage[version=3]{mhchem}
\usepackage[left=1.5cm, right=1.5cm, top=1.785cm, bottom=2.0cm]{geometry}
\usepackage{balance}
\usepackage{mathptmx}
\usepackage{sectsty}
\usepackage{graphicx} 
\usepackage{lastpage}
\usepackage[format=plain,justification=justified,singlelinecheck=false,font={stretch=1.125,small,sf},labelfont=bf,labelsep=space]{caption}
\usepackage{float}
\usepackage{fancyhdr}
\usepackage{fnpos}
\usepackage[english]{babel}
\addto{\captionsenglish}{%
}
\usepackage{array}
\usepackage{droidsans}
\usepackage{charter}
\usepackage[T1]{fontenc}
\usepackage[usenames,dvipsnames]{xcolor}
\usepackage{setspace}
\usepackage[compact]{titlesec}
\usepackage{hyperref}

\usepackage{epstopdf}

\definecolor{cream}{RGB}{222,217,201}
\begin{document}

\pagestyle{fancy}
\thispagestyle{plain}
\fancypagestyle{plain}{
\renewcommand{\headrulewidth}{0pt}
}

\makeFNbottom
\makeatletter
\renewcommand\LARGE{\@setfontsize\LARGE{15pt}{17}}
\renewcommand\Large{\@setfontsize\Large{12pt}{14}}
\renewcommand\large{\@setfontsize\large{10pt}{12}}
\renewcommand\footnotesize{\@setfontsize\footnotesize{7pt}{10}}
\makeatother

\renewcommand{\thefootnote}{\fnsymbol{footnote}}
\renewcommand\footnoterule{\vspace*{1pt}%
\color{cream}\hrule width 3.5in height 0.4pt \color{black}\vspace*{5pt}} 
\setcounter{secnumdepth}{5}

\makeatletter 
\renewcommand\@biblabel[1]{#1}            
\renewcommand\@makefntext[1]%
{\noindent\makebox[0pt][r]{\@thefnmark\,}#1}
\makeatother 
\renewcommand{\figurename}{\small{Fig.}~}
\sectionfont{\sffamily\Large}
\subsectionfont{\normalsize}
\subsubsectionfont{\bf}
\setstretch{1.125} 
\setlength{\skip\footins}{0.8cm}
\setlength{\footnotesep}{0.25cm}
\setlength{\jot}{10pt}
\titlespacing*{\section}{0pt}{4pt}{4pt}
\titlespacing*{\subsection}{0pt}{15pt}{1pt}

\fancyfoot{}
\fancyfoot[LO,RE]{\vspace{-7.1pt}\includegraphics[height=9pt]{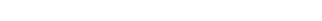}}
\fancyfoot[CO]{\vspace{-7.1pt}\hspace{13.2cm}\includegraphics{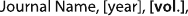}}
\fancyfoot[CE]{\vspace{-7.2pt}\hspace{-14.2cm}\includegraphics{head_foot/RF}}
\fancyfoot[RO]{\footnotesize{\sffamily{1--\pageref{LastPage} ~\textbar  \hspace{2pt}\thepage}}}
\fancyfoot[LE]{\footnotesize{\sffamily{\thepage~\textbar\hspace{3.45cm} 1--\pageref{LastPage}}}}
\fancyhead{}
\renewcommand{\headrulewidth}{0pt} 
\renewcommand{\footrulewidth}{0pt}
\setlength{\arrayrulewidth}{1pt}
\setlength{\columnsep}{6.5mm}
\setlength\bibsep{1pt}

\makeatletter 
\newlength{\figrulesep} 
\setlength{\figrulesep}{0.5\textfloatsep} 

\newcommand{\topfigrule}{\vspace*{-1pt}%
\noindent{\color{cream}\rule[-\figrulesep]{\columnwidth}{1.5pt}} }

\newcommand{\botfigrule}{\vspace*{-2pt}%
\noindent{\color{cream}\rule[\figrulesep]{\columnwidth}{1.5pt}} }

\newcommand{\dblfigrule}{\vspace*{-1pt}%
\noindent{\color{cream}\rule[-\figrulesep]{\textwidth}{1.5pt}} }

\makeatother

\twocolumn[
  \begin{@twocolumnfalse}
{
\includegraphics[width=18.5cm]{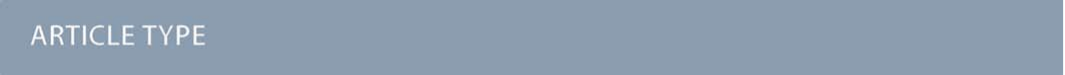}}\par
\vspace{1em}
\sffamily
\begin{tabular}{m{4.5cm} p{13.5cm} }

\includegraphics{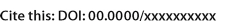} & \noindent\LARGE{\textbf{Impact of Disorder Dynamics and Multi-Domain Kinetics on the Sliding Ferroelectricity of CVD-Grown 3R-WSe$_2$ Bilayers}} \\
\vspace{0.3cm} & \vspace{0.3cm} \\

 & \noindent\large{Sourav Paul,\textit{$^{a}$} Prasenjit Ghosh,\textit{$^{a}$} Krishna Prasad Maity,\textit{$^{b}$} Vineet Pandey,\textit{$^{a}$} Abhijith M.B.,\textit{$^{a}$} Premananda Chatterjee,\textit{$^{c}$} Kenji Watanabe,\textit{$^{d}$} Takashi Taniguchi,\textit{$^{e}$} Nicholas R. Glavin,\textit{$^{f}$} Ajit K. Roy, \textit{$^{f}$} Atindra Nath Pal,\textit{$^{c}$}  and Vidya Kochat,$^{\ast}$\textit{$^{a}$} } \\

\includegraphics{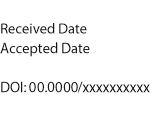} & \noindent\normalsize{Sliding ferroelectricity in van der Waals (vdW) layered systems has emerged as a promising route toward non-volatile nanoscale devices, where interlayer displacement in non-centrosymmetric bilayers generates an out-of-plane polarization. In particular, 3R-stacked bilayer transition metal dichalcogenides (TMDs) grown via chemical vapor deposition (CVD) have been shown to host such polarization due to broken inversion symmetry. However, a detailed investigation of the 2D ferroelectric (FE) properties of CVD-grown 2D films, particularly the role of intrinsic disorder, such as structural defects and domain structure, remains poorly understood. Here, we investigate the FE switching characteristics of CVD-grown 3R-stacked WSe$_2$ using a graphene-based ferroelectric field-effect transistor (graphene-FE-FET) architecture, where graphene serves as a highly sensitive probe of induced charge modulation due to polarization switching of FEs. We show that the growth-induced structural disorder significantly impacts polarization switching, while multi-domain kinetics governs the evolution of the FE response. These findings provide important insights into the design and optimization of FE devices based on vdW materials.}

\end{tabular}

 \end{@twocolumnfalse} \vspace{0.6cm}

  ]

\renewcommand*\rmdefault{bch}\normalfont\upshape
\rmfamily
\section*{}
\vspace{-1cm}


\footnotetext{\textit{$^{a}$~Materials Science Centre, Indian Institute of Technology, Kharagpur, West Bengal 721302, India. $^{\ast}$ E-mail: vidya@matsc.iitkgp.ac.in}}
\footnotetext{\textit{$^{b}$~Department of Physics, SRM University AP, Andhra Pradesh - 522240, India}}
\footnotetext{\textit{$^{c}$~S. N. Bose National Centre for Basic Sciences, Kolkata, West Bengal -700106, India}}
\footnotetext{\textit{$^{d}$~Research Center for Electronic and Optical Materials, National Institute for Materials Science, 1-1 Namiki, Tsukuba 305-0044, Japan}}
\footnotetext{\textit{$^{e}$~Research Center for Materials Nanoarchitectonics, National Institute for Materials Science, 1-1 Namiki, Tsukuba 305-0044, Japan}}
\footnotetext{\textit{$^{f}$~Air Force Research Laboratory, Wright-Patterson Air Force Base, Ohio 45433-7718, USA}}



\section{Introduction}
Ferroelectric (FE) materials have emerged as promising candidates for next-generation transistor gate dielectrics, owing to their ability to retain electrical polarization in the absence of an external gate voltage, thereby reducing gate leakage. The tunability of polarization under applied electric fields further enables their integration into a wide range of electronic and nanoelectronics applications, particularly in non-volatile memory technologies, such as FE random-access memory.~\cite{FE_application_science,FE_application_NatElect,FE_application_PiP,FE_appl_APL,FE_appl_NatNanotech,FE_application_AdvMat} In the realm of next-generation technology, two-dimensional (2D) FEs are expected to be highly desirable candidates for innovative device architectures for nonvolatile memory devices, neuromorphic computing, opto-synaptic applications.~\cite{FE_appl_NeuComEng,FE_appl_SmallMethod} In 2D systems, ferroelectricity arises from in-plane atomic displacements that overcome depolarization fields and break centrosymmetry, enabling stable polarization at reduced dimensionality. This switching behavior is energy-efficient and well-suited for integration into nanoscale devices.~\cite{perovskite_FE_science,perovskite_FE_nanoletter,perovskite_FE_jacs,perovskite_FE_dt,perovskite_FE_jms,HfO2_NatRevMat,HfO2_nanoletter} For examples,  CuInP$_2$S$_6$ (CIPS) exhibits robust out-of-plane ferroelectricity down to 4 nm at room temperature, driven by the displacement of Cu atoms relative to adjacent layers.~\cite{CIPS_acsnano,CIPS_ScAdv,CIPS_NatCom,CIPS_nanoletter} Similarly, $\alpha$-In$_2$Se$_3$ demonstrates both in-plane and out-of-plane ferroelectricity.~\cite{In2Se3_acsnano,In2Se3_AdvFunMat,In2Se3_AdvFunMat2,In2Se3_nanoletter,In2Se3_nanoletter2}. While earlier investigations using piezo force microscopy (PFM), scanning tunneling microscopy (STM), or transport measurements have indicated the FE properties of these materials at extremely thin dimensions, challenges remain in producing stable ultra-thin 2D films with out-of-plane polarization from these conventional 2D FE materials.

Beyond intrinsic mechanisms, a stable long-range FE ordering can also emerge in 2D van der Waals (vdW) semiconductors through interlayer sliding and twisting, a phenomenon known as sliding ferroelectricity. Relative lateral displacement by approximately one lattice constant, or the introduction of a twist angle between layers, breaks inversion symmetry and induces asymmetric interlayer charge transfer, giving rise to bistable polarization states separated by a low-energy barrier.~\cite{Sliding_FE_Theory_ACSNano,Sliding_FE_Theory_PRL,Sliding_FE_PRL,2D_FE_NatMat,Sliding_FE_PNAS,Sliding_FE_PFM_Sc} Recent studies on 2D ferroelectricity have revealed the emergence of sliding ferroelectricity in twisted bilayer systems, which generate moiré domains with opposite polarizations adjacent to one another. This includes twisted bilayer hBN and even parallel-stacked 2D TMDs such as WSe$_2$, MoSe$_2$, WS$_2$, MoS$_2$, and WTe$_2$.~\cite{FE_3RWSe2_AdMat,Sliding_FE_MoS2_NatElec,Sliding_FE_MoS2_NatCom,tTMD_FE_Sc,tTMD_FE_Nat_Nanotech} 

\begin{figure*}[h]
\centering
  \includegraphics[width=1\textwidth]{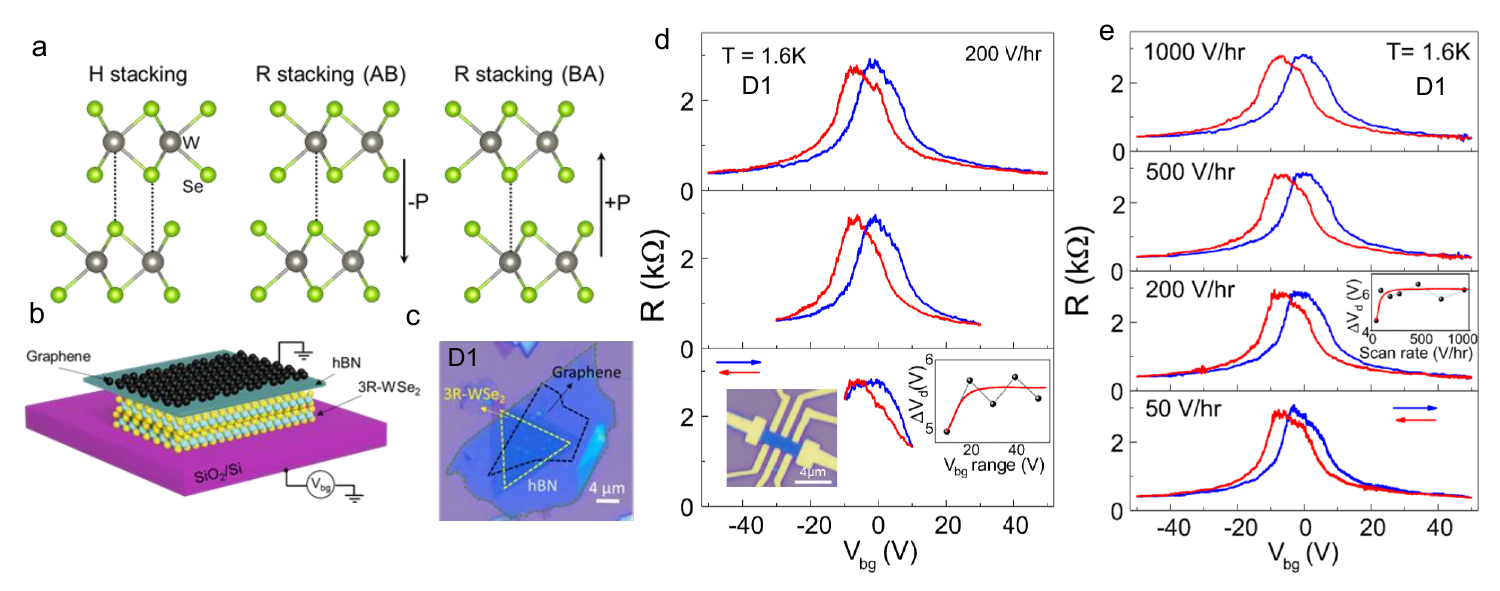}
  \caption{(a) Atomic arrangement of bilayer WSe$_2$ with different stacking configurations. While Hexagonal (H) stacking preserves inversion symmetry, a Rhombohedral (R) stacked WSe$_2$ bilayer with broken inversion symmetry exhibits out-of-plane polarization in opposite directions for the two different stacking sequences ($-\textbf{P}$ for AB and $+\textbf{P}$ for BA). (b) Schematic representation of Graphene/hBN/3R-WSe$_2$ heterostructure on SiO$_2$/Si substrate for graphene-FE-FET device. (c) Optical image of graphene/hBN/3R-WSe$_2$ heterostructure for device D1. (d) Resistance ($R$) of graphene (device D1) as a function of bottom gate voltage ($V_{bg}$) at temperature T = 1.6~K for different sweep ranges of V$_{bg}$ with scan rate 200~V/hr. Forward and backward $V_{bg}$ scans are represented by blue and red curves respectively. Left inset: Optical micrograph of device D1 in a Hall bar etched configuration. Right inset: Separation between the forward and backward scan gate voltage values corresponding to CNP ($\triangle V_d$) as a function of $V_{bg}$ scan range showing a saturation represented by the red line which is a guide to eye. (e) $R$ as a function of V$_{bg}$ for different scan rates for the device D1 at T= 1.6K for a fixed range of gate voltage. Inset: $\triangle V_d$ as a function of $V_{bg}$ scan rate showing saturation represented by the red line which is a guide to eye.} \label{Figure_1}
\end{figure*}

Monolayer (ML) WSe$_2$ exhibits semiconductor properties with broken inversion symmetry, resulting in distinctive features like strong piezoelectricity. For bilayer WSe$_2$, the 2H stacking represents the most stable configuration, characterized by a 180$^\circ$ rotation between the two layers, resulting in inversion symmetry. However, the two MLs with identical crystallographic configurations stacked without relative orientation are also energetically preferable for these vdW-layer TMDs. In the context of CVD growth of bilayer TMDs, the formation of 3R-stacked bilayers, in addition to 2H-stacked bilayers, is feasible under specific growth conditions. The atomic arrangement in the 3R stacking configuration disrupts the inversion symmetry at the interface, leading to an out-of-plane electric dipole moment with opposite directions in the non-centrosymmetric AB and BA stacking configurations, thereby facilitating interfacial charge transfer between the two layers as shown in fig.~\ref{Figure_1}a. Recent investigations of 2D ferroelectricity in 2D TMDs using PFM reveal that sliding ferroelectricity is observed in 3R-stacked bilayers, driven by domain-wall motion.~\cite{FE_3RWSe2_AdMat} Nonetheless, the extensive investigation into sliding ferroelectricity in CVD-grown TMDs, crucial for the fabrication of large-scale 2D memory devices, remains insufficiently explored. A comprehensive investigation into the effects of disorder and domain formation on sliding FE switching in CVD-grown TMD bilayers is still lacking. In this work, we investigate sliding ferroelectricity in CVD-grown 3R-stacked bilayer WSe$_2$, with particular emphasis on the temperature dependence of the switching behavior, probed via charge transport measurements in an adjacent graphene layer. We examine the impact of intrinsic disorder arising from growth-induced structural defects and multi-domain kinetics on ferroelectricity in 3R-TMD bilayers.

\section{Experiment}

Bilayer WSe$_2$ films were synthesized on c-plane sapphire substrates using the CVD technique (See supporting information section 1 (i)). CVD-grown 3R stacked bilayer WSe$_2$ films used for device fabrication were identified by optical contrast and Raman spectroscopy and the FE nature was confirmed through PFM studies.~\cite{Raman_3R_JAP} (See supporting information section 1 (ii, iii)) Graphene/hBN/3R-WSe$_2$ heterostructures (fig.~\ref{Figure_1}b) were fabricated by a modified 2D transfer method that integrates dry and wet transfer techniques (see the supporting information section 2 for details). Fig.~\ref{Figure_1}c shows the optical image of the heterostructure. Strong spin-orbit coupling, interface disorder, and structural defects in WSe$_2$ can induce significant charge inhomogeneity in the graphene channel, thus reducing charge carrier mobility and widening the graphene resistance vs. gate voltage ($R-V_{bg}$) characteristics.~\cite{FE_transport_tTMDs_NatNanotech} Therefore, we introduce hBN (thickness of $\sim$8-15~nm) as a spacer layer between graphene and WSe$_2$. The electrical connections to this graphene ferroelectric field effect transistor (Graphene-FE-FET) were fabricated in a Hall bar geometry through the process of reactive ion etching followed by thermal evaporation of Ti/Au (10 nm/60 nm). In this work, three Graphene-FE-FET devices (D1, D2, D3) are examined to investigate FE properties of 3R stacked bilayer WSe$_2$. The inset of fig.~\ref{Figure_1}d shows the optical image of the Hall bar channel for device D1 with length L = 4.1 $\mu$m and a width of W = 2.2 $\mu$m.

\begin{figure*}[h]
 \centering
 \includegraphics[width=1\textwidth]{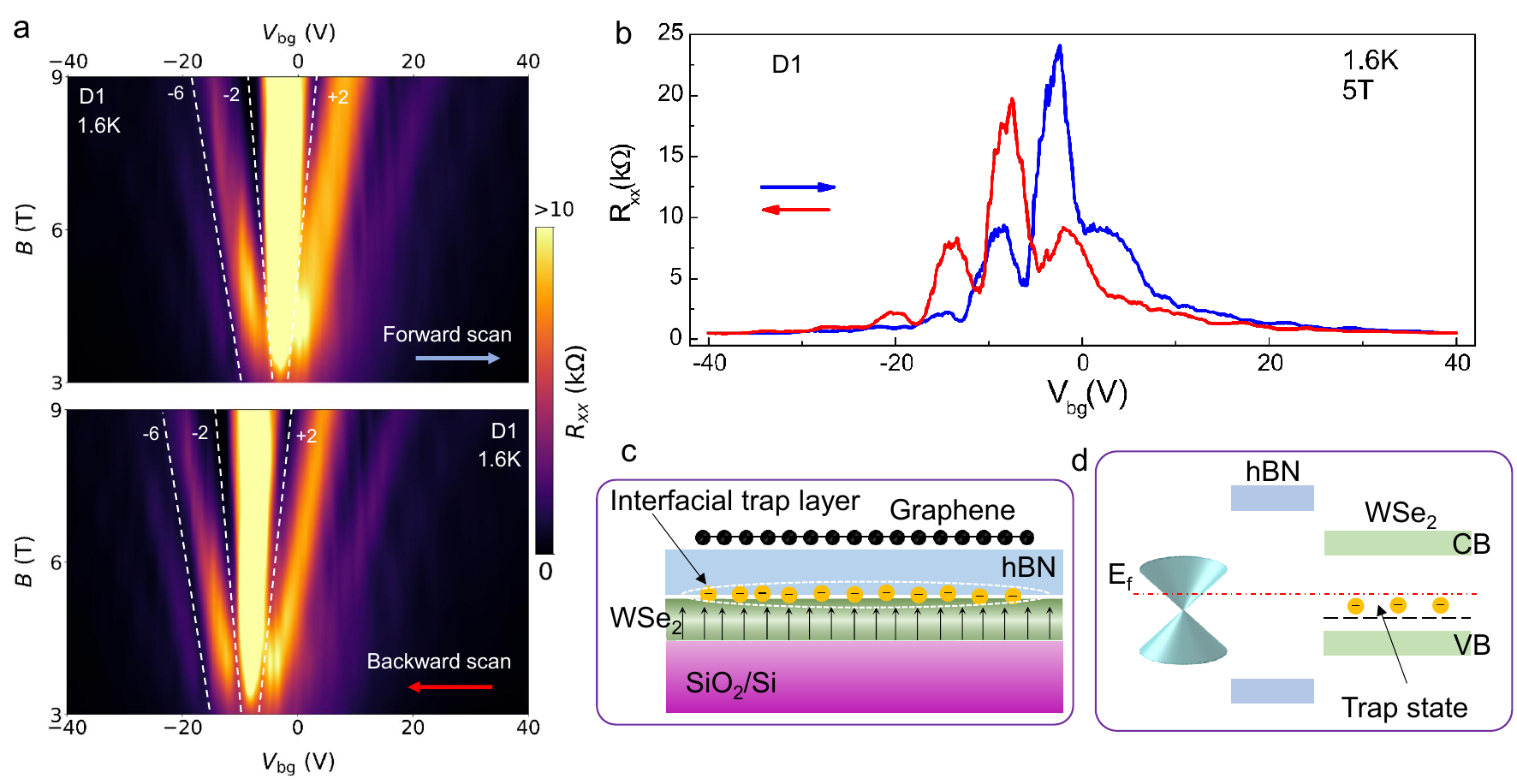}
 \caption{(a) The Landau fan diagram for device D1 at T= 1.6~K illustrates the longitudinal resistance, $R_{xx}$, plotted against the magnetic field on the vertical axis and the gate bias on the horizontal axis for both the forward (top) and backward (bottom) sweeps. Different filling factor values are marked in white dashed lines. (b) $R_{xx}$ as a function of $V_{bg}$ voltage sweep for both forward and backward scans at B = 5~T for device D1. (c) The schematic illustrates the formation of the interfacial charge layer at the hBN-WSe$_2$ interface. (d) A schematic of band diagram for graphene/hBN/3R-WSe$_2$ heterostructure shows the formation of localized defect states near the valence band (VB) maxima of WSe$_2$.}
 \label{Figure_2}
\end{figure*}

\section{Result and discussion}
The transfer characteristics of Graphene-FE-FET device, D1 obtained through a standard lock-in amplifier in a four-probe configuration, is shown in fig.~\ref{Figure_1}d, as a function of back gate voltage at temperature T = 1.6~K. The bottom gate voltage ($V_{bg}$) sweep with a voltage scan rate of 200 V/hr reveals a significant shift in the gate voltage values corresponding to the charge neutrality point (CNP) of graphene channel. This can be seen from the shift of the maximum resistance values of the $R-V_{bg}$ curve between the forward and backward scan directions of back gate leading to hysteretic behaviour. Here graphene serves as an effective electronic sensor, enabling precise estimation of the polarization of 3R-WSe$_2$. The hysteresis in the bottom gate scan in D1 arises due to FE polarization switching in 3R-WSe$_2$.~\cite{FE_transport_thBN_NatCom,FE_transport_thBN_Nature,FE_transport_tWSe2_NanoLetter,FE_transport_tTMDs_NatNanotech} The shift of the CNP resistance peak in the $V_{bg}$ - axis indicates an abrupt change in the charge density induced in graphene channel. A finite back gate voltage creates an electric field across the 3R-WSe$_2$ bilayer system, lowering the energy barrier required for inter-layer sliding and leads to a switching between the two stable stacking configurations (AB stacking to BA stacking). This causes the polarization to switch direction which modifies the net displacement field and the net induced charge density in the graphene channel changes from $(\frac{C_{tot}V_{bg}}{e} + \triangle n_p)$ to $(\frac{C_{tot}V_{bg}}{e} - \triangle n_p)$, where $C_{tot}$ is total bottom gate capacitance per unit area, given by $C_{tot}$ = $\frac{1}{\frac{1}{C_{SiO_2}} + \frac{1}{C_{hBN}} + \frac{1}{C_{WSe_2}}}$, $\triangle$n$_p$ is the induced charge carrier density due to the FE polarization in 3R-WSe$_2$ and $e$ is the electron charge.~\cite{FE_transport_thBN_Sc, FE_transport_thBN_Sc2, FE_transport_thBN_Nature, FE_transport_thBN_NatCom, FE_transport_tTMDs_NatNanotech} The dependence of the hysteretic transfer characteristics for device D1 on varying sweep voltage ranges at 1.6~K shows that there is no significant change in hysteresis as the gate voltage sweep range increases from 10 to 50~V, as shown in fig.~\ref{Figure_1}d, indicating a robust FE nature for the device D1. Typically, when the gate voltage range overcomes the coercive field of a single domain FE, the  hysteresis remains consistent or saturated.~\cite{mag_trap_nano_letter} For more clarity, we also performed transport characteristics analysis at various ramping rates of the gate voltage within a fixed voltage range of -50~V to +50~V. The FE hysteresis demonstrates minimal dependence on the gate voltage ramp rate as shown in fig.~\ref{Figure_1}e, indicating that the observed hysteresis in the transport characteristics of D1 can be attributed solely to the FE switching process of a single domain. At very slow scan rate of 50~V/hr, hysteresis width is smaller due to the quasi-static polarization switching under near-equilibrium conditions.
From the saturated value of the voltage shift of the CNP / Dirac point resistance between the forward and backward gate voltage scans ($\triangle V_d$) (indicated in the insets of fig.~\ref{Figure_1}d,e), $\triangle$n$_p$ can be estimated using the equation, 
\begin{equation} \label{eq:1}
  2\triangle n_p =  C_{tot} \times \frac{\triangle V_d}{e}  
\end{equation}
We obtain a value of ~$4.03 \times 10^{11}$ cm$^{-2}$ for $2\triangle$n$_p$ in D1 at T = 1.6~K. The magnitude of 2D polarization in 3R-WSe$_2$, $P_{2D}$, is estimated using the formula, 
\begin{equation} \label{eq:2}
    P_{2D}=e\triangle n_pd_{B}
\end{equation}
as $2.61 \times 10^{-12}~C/m$ at T = 1.6~K. Here, $d_{B}$ is the thickness of hBN. ~\cite{FE_transport_tWSe2_NanoLetter} 
\begin{figure*}[h]
 \centering
 \includegraphics[width=1\textwidth]{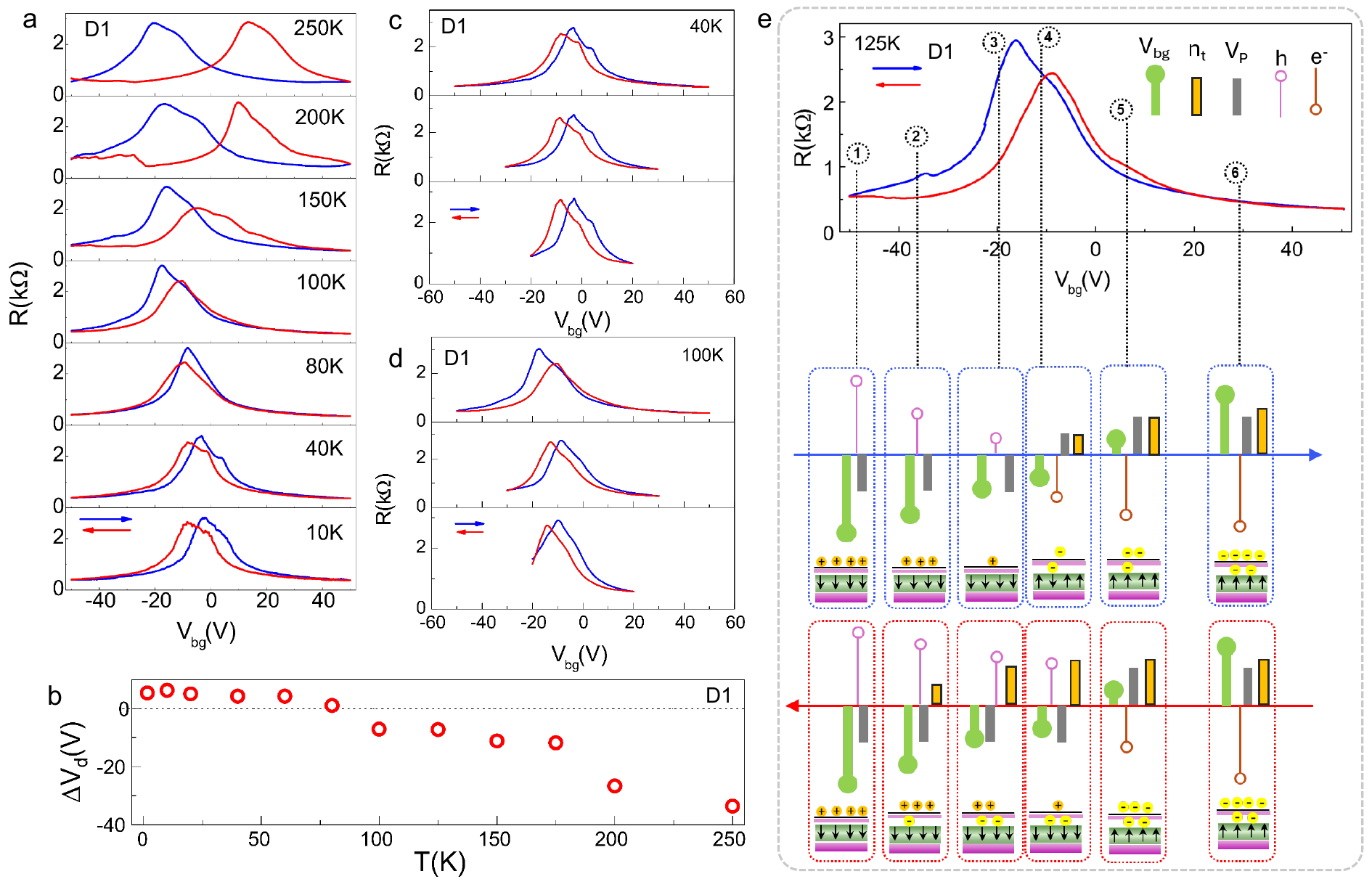}
 \caption{(a) Hysteresis and anti-hysteresis in transfer characteristics of device D1 as a function of V$_{bg}$ with scan rate 200~V/hr for different temperatures. (b) $\triangle V_d$ as a function of temperature. (c-d) Hysteresis, and anti-hysteresis observed in transfer characteristics for different sweep ranges of $V_{bg}$ with fixed scan rate of 200~V/hr at temperature (c) T = 40~K and (d) T = 100~K. (e) Illustration of trap-sensitive electrostatic model for $R-V_{bg}$ characteristics of D1 at 125~K.}
 \label{Figure_3}
\end{figure*}
    
The polarization switching in 3R-WSe$_2$ results in the hysteresis in the transfer characteristics of graphene and hence acts as a tunable knob for graphene's electronic spectrum. It was earlier reported that FE substrates can stabilize Quantum Hall effect (QHE) in graphene at higher temperatures and lower magnetic fields than pristine graphene.~\cite{FE_QHE_Com_Ph} FE polarization in 3R-WSe$_2$ shifts the Fermi level of graphene and alters the filling of Landau levels, whereby the QHE response inherits the hysteretic nature. The Landau fan diagram presented in fig.~\ref{Figure_2}a illustrates the evolution of longitudinal resistance (R$_{xx}$) as a function of magnetic field while tuning the gate voltage in the forward direction (top panel), while the data corresponding to the backward gate voltage sweep is displayed in the bottom panel. The white lines in the Landau fan correspond to the minima observed in $R_{xx}$, illustrating the Landau level filling factor, $\nu$, at values of 2, 6, and 10 corresponding to the integer quantum Hall sequence in ML graphene given by equation, $\nu$ = $4n+2$, where n is an integer. The $R_{xx}$ minima are clearly shifted depending on the sweep direction of the gate voltage, embedding memory effects into quantum transport. In fig.~\ref{Figure_2}b, the R$_{xx}$ measured at a constant magnetic field of 5~T in device D1 is plotted against the $V_{bg}$ sweep for both forward and backward scans, respectively. In addition to the hysteretic QHE, we also observe an asymmetry in the $R_{xx}$ evolution with Landau level formation in the hole-doped and electron-doped regions.  QHE is a highly sensitive probe of disorder present in the system. The blurring of $R_{xx}$ minima and mixing of Hall resistance with $R_{xx}$ is observed in the electron-doped QH regime, which suggests Landau level broadening and loss of clean quantization arising from a disordered potential landscape seen by carriers in graphene.~\cite{mag_trap_nano_letter} We attribute this asymmetric QHE to the remote Coulomb disorder arising from the well-established Se vacancies in WSe$_2$ created during the CVD growth process, which form localized defect states close to the valence band maxima of WSe$_2$ as shown in fig.~\ref{Figure_2}d. As the gate voltage is tuned to positive side, these defect states are populated with electrons, leading to the formation of an interfacial charge layer (ICL) as shown in fig.~\ref{Figure_2}c. Thus the occupied Se vacancies form a fluctuating screening layer based on the defect distribution in the 3R-WSe$_2$ crystal and modify the effective potential landscape seen by the electrons in graphene. Sweeping the gate voltage to negative side lowers the Fermi level below the defect level and the ICL vanishes due to the empty defect state. The disordered potential landscape mixes the Hall and longitudinal conductance values on the positive gate voltage side, while the plateaus are stabilized with well-defined $R_{xx}$ minima in the hole-doped transport regime in graphene. Thus QHE in graphene serves as a direct indicator of disorder nature and strength in 3R-WSe$_2$.

\begin{figure*} [h]
 \centering
 \includegraphics[width=1\textwidth]{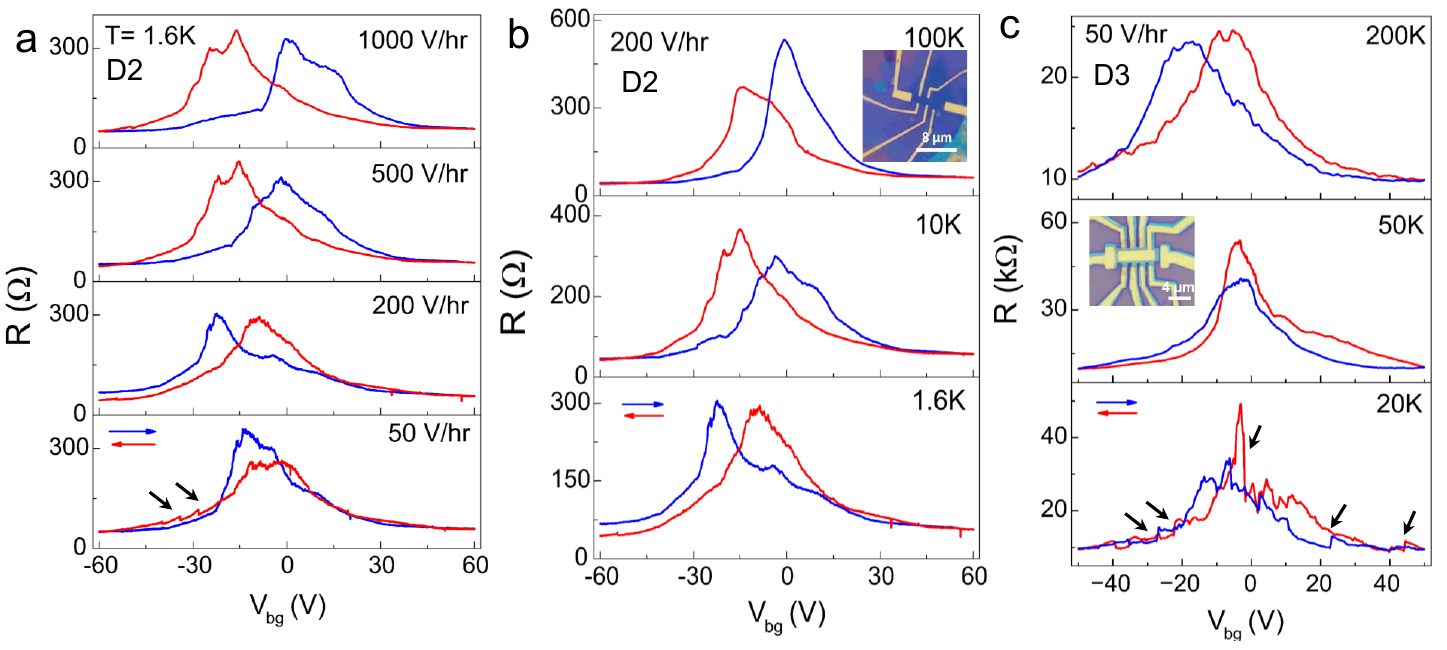}
 \caption{(a) $R-V_{bg}$ characteristics as a function of V$_{bg}$ at 1.6~K for different gate voltage scan rates for device D2. (b,c) Transfer characteristics for device (b) D2 and (c) D3 at different temperatures. Inset: Optical image of devices D2 and D3, respectively. Black arrow marks in fig. (a) and (c) indicate the sudden resistances jumps in graphene transport due to the relaxation of domains in 3R-WSe$_2$.}\label{Figure_4}
\end{figure*}

As we have already established, the ML graphene transport is a direct sensor of FE polarization in 3R-WSe$_2$ at low temperatures. Understanding the evolution of hysteresis with temperature is essential to demonstrate reliable non-volatile memory devices. Fig.~\ref{Figure_3}a shows the temperature-dependent $R-V_{bg}$ characteristics of device D1, obtained at a fixed sweep rate of 200 V/hr for forward and backward gate sweep scans. The striking observation here is the lowering of hysteresis (or $\triangle V_d$) as temperature increases from 10 to 80~K, above which an interesting anti-hysteresis behavior emerges, which consistently increases with further increase in temperature to 250~K. The hysteresis to anti-hysteresis transport behaviour develops at $T\sim80$~K as can be observed from the sign change from positive to negative $\triangle V_d$ in fig.~\ref{Figure_3}b. We next examined the transfer characteristics in D1 as a function of $V_{bg}$ sweep range at a fixed scan rate of 200 V/hr, in order to draw comparison at two representative temperatures: 40 K, below the hysteresis–anti‑hysteresis transition, and 100 K, above this transition. Compared to the sweeps at 1.6~K where a saturation of $\triangle V_d$ was observed with increase in $V_{bg}$ sweep range, (fig.~\ref{Figure_1}d) the $\triangle V_d$ at 40~K showed a decrease, indicating reduced hysteresis with increase in $V_{bg}$ sweep range, as can be seen from fig.~\ref{Figure_3}c. On the contrary, at 100~K, a hysteresis to anti-hysteresis dominated transport was observed with an increase in the $V_{bg}$ sweep range which is depicted in fig.~\ref{Figure_3}d. This indicates the presence of a thermally activated process competing with FE polarization switching at elevated temperatures, that modulates the hysteretic to anti-hysteretic transport in graphene.  In general, the anti-hysteresis phenomenon in graphene/FE structures have been attributed to the dominant role of charge trapping/de-trapping processes due to an ICL from adsorbates, which outweighs the effect of the competing FE switching process.~\cite{Charge_trap_JETP,Anti_hys_FE_NanoLetter} Our observation of anti-hysteresis is also strikingly different from earlier reports on similar heterostructure devices fabricated on exfoliated 3R-stacked TMDs, where the hysteresis due to polarization switching was evident even at room temperature.~\cite{FE_transport_tTMDs_NatNanotech} This again points to the role of Se vacancies created in 3R-WSe$_2$ during the CVD growth process, whose densities are significantly higher in comparison to exfoliated WSe$_2$. The Se vacancies introduce localized defect states that act as charge traps, creating an ICL in graphene/h‑BN/3R‑WSe$_2$ heterostructure devices, as shown in fig.~\ref{Figure_2}c.~\cite{interfacial_charge_ACSNano} 

The origin of anti-hysteresis can be understood based on a trap-sensitive electrostatic model presented in fig.~\ref{Figure_3}e. This provides a framework to describe how the applied gate voltage plus the FE polarization in 3R-WSe$_2$ and the defect states from Se vacancies jointly shape the electrostatic environment of graphene. When a gate voltage is applied, these traps dynamically capture or release carriers, forming an ICL that screens the FE polarization field. At low temperatures, FE polarization switching dominates, producing conventional hysteresis in transport. However, as temperature increases, the trap dynamics become thermally activated and respond faster than the FE dipoles. This competition means that instead of reinforcing the polarization‑induced field, the trapped charges can reduce or even invert the effective electric field seen by the graphene channel, leading to anti‑hysteresis. Thus, the ICL arising from Se vacancies effectively counteracts and eventually overrides FE switching, flipping the hysteresis loop orientation in graphene transport. The graphene carrier density ($n_G$) is determined by the net electrostatic field, which is the sum of the applied gate voltage ($V_{bg}$), FE polarization in 3R-WSe$_2$ ($P_{2D}$), and the trap contribution $n_{t}$, which changes dynamically with applied gate voltage and temperature. 
\begin{equation} \label{eq:3}
    n_{G}=\frac{C_{tot}(V_{bg}-V_{d})}{e} + \frac{P_{2D}}{e}-n_{t}(V_{bg},T)
\end{equation}
We now describe the electrostatic model using the anti-hysteretic transfer characteristics of graphene channel obtained at 125~K as shown in fig.~\ref{Figure_3}e top panel. The bottom panel describes an effective field balance diagram depicting the magnitudes of the individual contributions from the above mentioned voltage parameters and the resultant magnitude of the charge carrier density electrostatically induced in graphene at different applied gate voltages. At the high negative gate voltages (points 1, 2, 3 of blue curve), the contributions from $V_{bg}$ and FE polarization, $V_P$, add up, thereby tuning the Fermi level in graphene to the valence band or hole-doped regime. As the gate voltage increases (towards 0~V) in the forward sweep, the AB and BA domains become energetically equivalent and at point 4 of the blue curve BA domains dominate leading to net positive $V_{P}$. This is evident from the observation of double resistance peak features in $R-V_{bg}$.~\cite{FE_transport_tTMDs_NatNanotech} At the same time, an ICL also forms due to the thermal activation of traps, which screens the $V_{bg}$. The combined effect of $+V_{P}$ and the ICL leads to electron doping in graphene by shifting the $E_F$ to the conduction band. At higher $+V_{bg}$, the electron density in graphene increases as polarization completely switches forming a single domain which adds to the applied $V_{bg}$, as seen at point 5 of the blue curve. With further increase in $+V_{bg}$, the $E_F$ in 3R-WSe$_2$ passes above the trap states, populating them with electrons and strengthens the ICL as shown at point 6 on the blue curve. If the thermal activation of trap states are ignored, then only the effect of FE polarization switching is added to the applied $V_{bg}$, resulting in hysteresis as observed at low temperatures. However, with increase in temperature, the traps also re-capture electrons with a net effect of slower detrapping rate due to the competition between carrier release and capture. This results in an overall stretching out of the detrapping process, even though $E_F$ is tuned by applied gate voltage to lie below the trap level energies. Due to this, the ICL persists even at $-V_{bg}$ values as shown in the schematics corresponding to the red curve. In the reverse gate voltage sweep denoted by the red curve, the screening of the applied $V_{bg}$ by this ICL reduces the electron density induced in graphene channel, leading to higher resistance at point 5 compared to the forward sweep. As $V_{bg}$ is tuned to negative values, the FE polarization switches direction, but the sluggish detrapping dynamics at higher temperatures screen the FE polarization, thereby reducing the effective field seen by graphene, leading to anti-hysteresis as seen while traversing the points in the red curve from 4 to 1. Above the transition temperature, corresponding to hysteresis to anti-hysteresis behavior, a more pronounced shift in the resistance peak position at CNP is observed between forward and reverse gate scans as seen in fig.~\ref{Figure_3}a from 150 to 250~K. This indicates the higher populations of trap states, which significantly increases the electron doping in graphene, shifting the CNP towards larger $-V_{bg}$ in the forward sweep scan. During the reverse gate sweep, the effective detrapping rate is drastically reduced, shifting the CNP to larger $+V_{bg}$ values enhancing the anti-hysteresis, as modelled using trap-assisted electrostatic fields. The increased remote Coulomb disorder also broadens the $R-V_{bg}$ curves at higher temperatures.

Till now, we have discussed the polarization switching and activated trap dynamics for a single domain FE 3R-WSe$_2$. In such a case, the polarization switching is sharp and is governed by the intrinsic coercive field, yielding stable hysteresis loops at low temperatures. The obvious next question is what happens in the case of a multi-domain FE 3R-WSe$_2$-based graphene FE-FET and how to identify the multi-domain FE behaviour from transport characteristics. In fig.~\ref{Figure_4}, we describe the polarization switching dynamics and disorder effects in multi-domain FE devices, D2 and D3. Contrary to the sweep-rate independent hysteresis loops observed in D1 at low temperatures (fig.~\ref{Figure_1}e), D2 displayed intriguing anti-hysteresis characteristics at slow sweep rate, which transitioned to hysteretic behaviour on increasing the sweep rate as shown in fig~\ref{Figure_4}a. At low temperatures, carrier trapping/detrapping is negligible, and the anti-hysteresis observed under slow gate sweeps arises from FE domain relaxation rather than trap dynamics. In a multi-domain FE, different domains switch at slightly different coercive fields. At slow sweeps, domain relaxation and partial back-switching reduce the net polarization and could produce an apparent anti-hysteresis. The relaxation of domains is evident from the sharp jumps (indicated by black arrows in fig~\ref{Figure_4}a) observed in the $R-V_{bg}$ curve obtained at a sweep rate of 50~V/hr, confirming the multi-domain hypothesis. At fast sweep rates, these domains are driven to switch abruptly, arresting the relaxation to yield conventional hysteresis. The fig~\ref{Figure_4}b describes the evolution of the forward and backward $R-V_{bg}$ scans of D2 at a fixed rate of 200~V/hr with temperature. The anti-hysteresis observed at 1.6~K evolves quickly to hysteretic behaviour as temperature is increased to 10~K and remains so even at higher temperatures. The domain wall mobility increases with temperature, and polarization switching dominates even at slower sweep rates, restoring hysteresis. This sharp crossover at 10~K highlights the strong temperature sensitivity of domain wall dynamics in van der Waals FEs. Contrary to the device D1, the observation of conventional hysteresis at 100~K in D2, without any significant shift in $\triangle V_d$, suggests lower trap density in D2 and negligible effects of trap dynamics on polarization switching. Further evidence is obtained from the QH measurements in D2, where the Landau levels in the electron-doped regime of graphene are better developed compared to D1, indicating lesser Coulomb disorder from ICL due to Se vacancy trap states. (See Supporting information fig. S11) Finally we investigated the combined influence of Se vacancies and multi-domain structure in 3R-WSe$_2$, as illustrated in fig~\ref{Figure_4}c for the 2-probe device (D3) with a device configuration similar to D1 and D2. At lower temperatures of $\sim$20~K, device D3 exhibits multiple resistance jumps and anti-hysteresis under a slow gate sweep rate of 50~V/hr signaling domain relaxation. With increase in temperature, the resistance jumps vanish; however, anti-hysteresis persists even upto higher temperatures of $\sim$200~K. This behaviour indicates that activated trap dynamics arising from the larger density of Se vacancies obscure the effects of intrinsic polarization switching in disordered multi-domain 3R-WSe$_2$ bilayer films. The investigation of sliding FEs using transport measurements opens up numerous opportunities for the design of wafer-scale non-volatile computing in memory devices utilizing 2D vdW FEs.  

\section*{Conclusions}
In conclusion, here we employed graphene as a sensitive probe to unravel the FE switching behavior of R-stacked WSe$_2$ bilayers with broken inversion symmetry. Through systematic investigations of the disorder dynamics and multi-domain effects in 3R-WSe$_2$, we demonstrated the detrimental influence of intrinsic disorder in WSe$_2$ on the polarization switching process. Our studies of domain relaxation kinetics in multi-domain 3R-WSe$_2$ bilayers provide valuable insights into their sustainability for applications in non-volatile computing. Collectively, these findings establish the CVD-grown TMDs as a distinctive platform for exploring sliding ferroelectricity, towards advanced 2D memory devices and neuromorphic architectures.

\section*{Author contributions}
S.P., K.P.M., and V.K. conceived the project. S.P. fabricated all heterostructures and devices with support from P.G., V.P., P.C., and A.N.P.. A.M.B. grew the 3R- WSe$_2$ samples. S.P., P.G., and A.M.B. performed other characterizations. S.P. performed all transport measurements and data analysis. K.W., and T.T. supplied the hBN crystals. A.N.P., A.K.R., and N.G. helped with discussions. S.P., V.K., and K.P.M. wrote the manuscript with inputs from all other authors.

\section*{Conflicts of interest}
There are no conflicts to declare.

\section*{Data availability}
The data supporting this article have been included as part of the supplementary information (SI). Supplementary information: detailed CVD growth for WSe$_2$ 2D film, detailed fabrication method of graphene/hBN/3R-WSe$_2$, additional transport measurements data for all devices.

\section*{Acknowledgements}
We acknowledge the facilities extended for the project by Prof. C. S. Tiwary and the useful discussions with him. S.P., P.G., V.P., A.M.B., and V.K. acknowledge funding support from the DST-Nanomission program of the Department of Science and Technology, Government of India (DST/NM/TUE/QM-1/2019), SPM facility at the Materials Science Centre, and the STEP facility, IIT Kharagpur. This material is based upon work supported by the Air Force Office of Scientific Research under award numbers FA2386-24-1-4093 and FA2386-26-1-4002. P.C. and A.N.P. acknowledge the Thematic Unit of Excellence on Nanodevice Technology (Grant No. SR/NM/NS-09/2011) and the Technical Research Centre (TRC) Instrument facilities of S. N. Bose National Centre for Basic Sciences, established under the TRC project of Department of Science and Technology (DST), Govt. of India. A.N.P. acknowledges DST Nano Mission: DST/NM/TUE/QM-10/2019. K.P.M thanks the Anusandhan National Research Foundation (ANRF), India, for financial support under the National Post-Doctoral Fellowship (NPDF) scheme [File Number: PDF/2023/001249]. K.W. and T.T. acknowledge support from the JSPS KAKENHI (Grants No. 21H05233 and No. 23H02052), the CREST (JPMJCR24A5), JST, and World Premier International Research Center Initiative (WPI), MEXT, Japan.



\balance


\bibliography{rsc} 
\bibliographystyle{rsc} 
\end{document}